# $B_{63}$: the most stable bilayer structure with dual aromaticity

Jinhuang Chen,[1] Rui Liao,[1] Linwei Sai,[2] Xue Wu[1] Jijun Zhao[3]

[1] State Key Laboratory of Metastable Materials Science & Technology and Key Laboratory for Microstructural Material Physics of Hebei Province, School of Science, Yanshan University, Qinhuangdao 066004, China

[2] School of Science, Hohai University, Changzhou 213022, China

[3] Key Laboratory of Materials Modification by Laser, Ion and Electron Beams (Dalian University of Technology), Ministry of Education, Dalian 116024, China

**Abstract**

The emergence of the first bilayer $B_{48}$, which has been both theoretically predicted and experimentally observed, as well as the recent experimental synthesis of bilayer borophene on Ag and Cu, has generated tremendous curiosity in the bilayer structure of boron clusters. However, the connection between the bilayer cluster and the bilayer borophene remains unknown. By combining a genetic algorithm and density functional theory calculations, a global search for the low-energy structures of $B_{63}$ clusters was conducted, revealing that the $C_s$ bilayer structure with three interlayer B–B bonds was the most stable bilayer structure. This structure was further examined in terms of its structural stability, chemical bonding, and aromaticity. Interestingly, the interlayer bonds exhibited electronegativity and robust aromaticity. Furthermore, the double aromaticity stemmed from diatropic currents originating from virtual translational transitions at both the σ and π electrons. This new boron bilayer is anticipated to enrich the concept of double aromaticity and serve as a valuable precursor for bilayer borophene.

**Introduction**

As a prototypical electron-deficient element adjacent to carbon, boron clusters display unique size-selected geometric structures and complex multi-center two-electron bonding modes.[1,2] In order to create a potential bottom-up approach from boron clusters to boron nanomaterials, an in-depth comprehension of the structures and properties of boron clusters is essential.

In the past two decades, photoelectron spectroscopy (PES) experiments combined with density functional theory (DFT) calculations have been used to accurately identify boron clusters.[3-8] Small-sized $B_n$ anions tend to form planar structures, and then transform to convex quasi-planar structures due to the peripheral bonds being shorter than the internal ones.[9] With size increasing, these quasi-planar structures often have pentagonal or hexagonal holes, which are a result of the electron-deficient internal atoms.[8,10] This vacancy is necessary to complete the σ electron system of the triangular boron lattice and stabilize the structure.[5,11] In particular, the $C_{6v}$ $B_{36}$ cluster with a central perfect hexagonal hole is a very stable quasi-planar structure that can be seen as the precursor to the α-sheet,[5] and has been successfully synthesized on a Ag(111) substrate.[12]

Compared to anionic states, neutral boron clusters do not typically display a planar structural pattern. For instance, Kiran et al. proposed a double-ring tubular structure as the ground state at size $n$ = 20,[4] which can be viewed as a rolled-up version of a quasi-planar structure. Subsequently, Tian et al. demonstrated that triple-ring tubular structures have greater binding energies than double-ring tubular structures at increased sizes, but do not evolve further into four-ring tubular structures.[13] Additionally, our research has confirmed that the most significant tubular structure is present at $n$ = 45 with three-ring,[14] and the specific interlaced π bonds between the internal and external rings play a fundamental role in structural stability. Moreover, the external rings with strong localized σ bonds have slightly smaller diameters than the internal ring, resembling the peripheral bond of a quasi-planar structure. Similarly, it is reasonable to suggest that the cage-like structural pattern is the inevitable consequence of increased curvature and shortened peripheral bonds. Zhao et al. have revealed the smallest hollow cage for $B_{28}$,[15] which was later confirmed in experiments.[16] Zhai et al. proposed the largest and most stable fullerene cage, $B_{40}$ borospherene, consisting of interwoven double-chain boron ribbons with hexagonal and heptagonal holes on the surface.[6] In a similar way, the quasi-planar and cage structural patterns also alleviate the electron-excessive of boron atoms via vacancies. However, hollow cages cannot continue to increase in size without an internal core for support. For instance, the well-known fullerene-like $B_{80}$ buckyball[17] was found to be less energetically stable, subsequently collapsing to a core-shell structure with lower energy and more robust aromaticity.[18] Recently, our group has demonstrated that the core-shell evolved from $B_{46}$ with a $B_4$-core to a complete $B_{96}$ with an icosahedral $B_{12}$-core,[19,20] and a $B_{12}$-core half-covered $B_{58}$ bridges the gap,[21] suggesting that the core-shell pattern developed from several core-atoms to an icosahedral $B_{12}$, with the shell gradually covering the $B_{12}$-core.[22,23] It is worth noting that the multi-$B_{12}$ core-shell structural pattern maybe the tendency for large-size boron clusters.[24] The high stability of the core-shell structural pattern is attributed to its superatomic electronic configuration and spherical aromaticity.

A novel bilayer structure has been observed to begin emerging at $n = 48$,[7,19] consisting of two monolayers held together by strong interlayer covalent bonds, which may be the result of the collapse of hollow cage motifs. This interlayer bond strengthens the electrostatic interactions between the layers, resulting in an increased interlayer coupling strength and stability from two to four as the size increases. Li's group has since predicted a range of bilayer structures ($n = 50−74$).[25-27] Furthermore, this bilayer structure is thought to be a precursor of the two-dimensional bilayer boron allotrope,[28-30] the bilayer boronphene were recently synthesized on Cu(111) and Ag(111) substrates[31-33] and found to possess remarkable conductivity and greater stability than the monolayer one.[34] Despite the extensive size range, the most stable bilayer structure has yet not to be determined, and the association between boron bilayer and bilayer boronphene is unclearly.

Taking the aforementioned questions into consideration, we evaluated the ground state structures of $B_{48}−B_{72}$ clusters and conducted a comprehensive theoretical investigation of $B_{63}$ clusters. Our founding revealed that the $B_{63}$ is the most stable bilayer boron nanocluster with dual aromaticity, thus making it an ideal structural motif for bilayer borophene.

**Method**

Extensive searches of the global minima of $B_{63}$ were conducted using a comprehensive genetic algorithm (CGA) code combined with density functional theory (DFT) calculations.[35] The screened child clusters were relaxed using the Perdew-Burke-Enzerhof functional and the double numerical d-polarization function basis within the generalized gradient approximation (GGA),[36,37] as implemented in the DMol$^3$ program.[38] The efficacy and efficiency of this CGA-DFT scheme has been validated in prior studies on pristine boron clusters.[14,19,21] Further information on CGA can be found in a review article.[35] The thirty lower-lying isomers were meticulously re-optimized using the (U)TPSSh[39] method with the 6-311+G(d) basis set.[40] Vibrational analysis was conducted to ensure that the obtained isomers were true minima without imaginary frequency. The average binding energies per atom ($E_b$), second-order energy differences ($\Delta^2 E$) and HOMO-LUMO gaps (split into $\alpha$ and $\beta$) were used to evaluate the energetic stability of $B_{48}$−$B_{72}$ clusters. All calculations were performed using the Gaussian16 package.[41]

The first-principle molecular dynamics (FPMD) simulations of the $B_{63}$ (I) in the NVT ensemble at 800K and 1000K were conducted for 8 ps with a time step of 1 fs, in order to assess their thermal stability. Chemical bonding was evaluated using the electron localization function (ELF)[42] and the adaptive natural density partitioning (AdNDP) approaches.[43] Nucleus-independent chemical shifts (NICS)[44,45] were carried out to evaluate the aromaticity, while the atomic dipole moment corrected Hirshfeld (ADCH) method was applied to calculate the atomic charge distribution.[46] Wave function analyses were generated using the Multiwfn 3.8(dev) code,[47] while anisotropy of the current induced density (ACID) was employed to study the aromaticity,[48] realized through the ACID code,[49] and the map was generated with POV-Ray render. Additionally, the gauge-including magnetically induced currents (GIMIC) analysis[50] was incorporated, via the GIMIC code.[51]

**Results and Discussions**

The low-lying isomers of the $B_{63}$ cluster located within 5 eV are presented in Fig. 1 and Fig. S1 of the Electronic Supplementary Information. The global minima of $B_{63}$ with $C_s$ symmetry, defined as $B_{63}$ (I), consists of two identical quasi-planar layers of $B_{27}$, connected by three interlayer bonds in the center and three $B_3$ chains on the waist, leaving three lateral windows with an alternating distribution. This structure is 1.09 eV lower in energy than the metastable $C_2$ $B_{63}$ bilayer structure with four interlayer bonds at the TPSSh/6-311+G(d) level. Additionally, seven out of the ten lower-lying isomers of $B_{63}$ are bilayer structures, with the tubular, core-shell, and quasi-planar isomers being higher in energy than $B_{63}$ (I) by 1.22, 1.67 and 2.36 eV, respectively. Furthermore, the bilayer pattern is also predominant in the size range of $n$ = 48−72, as illustrated in Fig. S2 of the Electronic Supplementary Information.

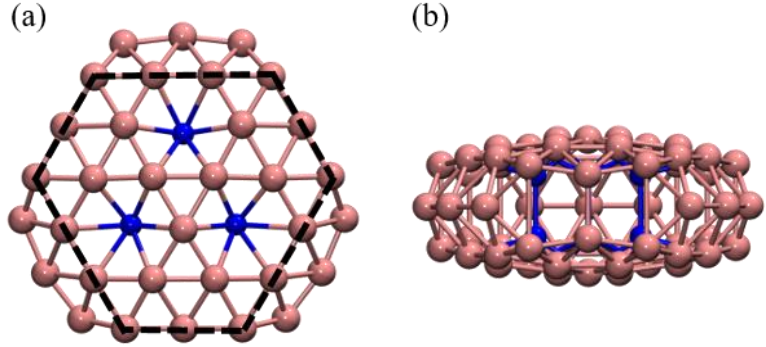

**Fig. 1** The ground-state structure of $C_s$ B$_{63}$ (I): (a) top view and (b) side view. Atoms buckled inward are highlighted in blue. The bilayer B$_{54}$ hexagonal prism is marked by black dashed line.

The stability of the B$_{63}$ (I) bilayer was further explored by calculating the binding energy ($E_b$), second-order energy difference ($\Delta^2 E$), and HOMO-LUMO gap ($E_{HL}$) of the ground states of even-sized B$_n$ clusters ($n$ = 48−72), as presented in Table 1 and Fig. 2. Generally, $E_b$ increased with cluster size, with a particularly notable peak at B$_{63}$. Additionally, the $\Delta^2 E$ of B$_{63}$ also exhibited a significant peak, suggesting its greater stability in comparison to other clusters in this size range. Furthermore, the $E_{HL}$ for the $\alpha$ electron of B$_{63}$ was relatively high at 1.00 eV (0.45 eV for the $\beta$ electron), demonstrating its strong electronic stability.

The structural evolution and stability of bilayer pattern have been studied (Table S1 and Fig. S3). Geometrically, the number of interlayer bonds increases with cluster size from two to four, and the average length of interlayer bonds initially increases and then decreases with cluster size. It is noteworthy that the largest size bilayer with three interlayer bonds ($C_s$ B$_{63}$ (I)) has the longest average interlayer bond length (1.716 Å), while the bilayer structures with two or four interlayer bonds have shorter average interlayer bond lengths. This value is comparable to the interlayer bond length of the bilayer-$\alpha$ borophene synthesized on Ag(111) substrate (1.8 Å),[31] and the freestanding $v_{1/12}$ bilayer borophene (1.73 Å) and bilayer-$\alpha^+$ borophene B$_{22}$ (1.76 Å).[30,33] Therefore, it is reasonable to suggest that $C_s$ B$_{63}$ (I) is the most suitable precursor for several bilayer borophenes.

To assess the thermodynamic stability of B$_{63}$ (I), FPMD simulations were carried out at 800 and 1000 K. The average root-mean-square deviation (RMSD) was 0.19 Å and 0.24 Å, respectively, demonstrating that all atoms remained close to their equilibrium positions. Furthermore, the energy fluctuated slightly, indicating its remarkable thermal stability at 1000 K (refer to Fig. S4).

**Table 1** The binding energy ($E_b$ in eV per atom), second-order energy difference ($\Delta^2 E$ in eV), HOMO-LUMO gap ($E_{HL}$ in eV), number of interlayer B−B bonds ($N_{B-B}$), and average bond length of interlayer B−B bonds ($d_{B-B}$ in Å) of the bilayer structure for the ground states of $B_n$ ($n$ = 48−72) clusters. The correspond structures are depicted in Fig. S2 of the Electronic Supplementary Information.

| Size $n$ | Structure | Symmetry | $N_{B-B}$ | $d_{B-B}$ | $E_b$ | $\Delta^2 E$ | $E_{HL}$ | Ref |
|---|---|---|---|---|---|---|---|---|
| 48 | bilayer | $D_{2h}$ | 2 | 1.656 | 5.205 | 0.865 | 0.93 | Ref.20 |
| 50 | bilayer | $C_1$ | 2 | 1.675 | 5.208 | –0.152 | 1.16 | Ref.27 |
| 52 | bilayer | $C_{2h}$ | 2 | 1.681 | 5.213 | –0.491 | 1.15 | Ref.27 |
| 54 | bilayer | $C_2$ | 3 | 1.683 | 5.227 | 0.716 | 1.39 | Ref.28 |
| 56 | core-shell | $C_{3v}$ | / | / | 5.228 | –0.378 | 1.00 | Ref.22 |
| 58 | core-shell | $C_{3v}$ | / | / | 5.234 | –0.534 | 0.79 | Ref.22 |
| 60 | bilayer | $C_s$ | 3 | 1.702 | 5.250 | 1.308 | 0.60 | Ref.22 |
| 62 | bilayer | $C_s$ | 3 | 1.715 | 5.243 | –0.659 | 0.65 | Ref.22 |
| 63 | bilayer | $C_s$ | 3 | 1.716 | 5.256 | 1.473 | 1.00($\alpha$),0.45($\beta$) | this work |
| 64 | core-shell | $C_2$ | / | / | 5.247 | –0.102 | 0.81 | Ref.22 |
| 66 | bilayer | $D_2$ | 4 | 1.686 | 5.252 | –0.186 | 0.76 | Ref.29 |
| 68 | bilayer | $D_2$ | 4 | 1.697 | 5.259 | 0.001 | 0.86 | Ref.29 |
| 70 | bilayer | $C_1$ | 4 | 1.690 | 5.267 | 0.211 | 0.78 | Ref.29 |
| 72 | bilayer | $C_i$ | 4 | 1.684 | 5.270 | 0.245 | 0.90 | Ref.29 |

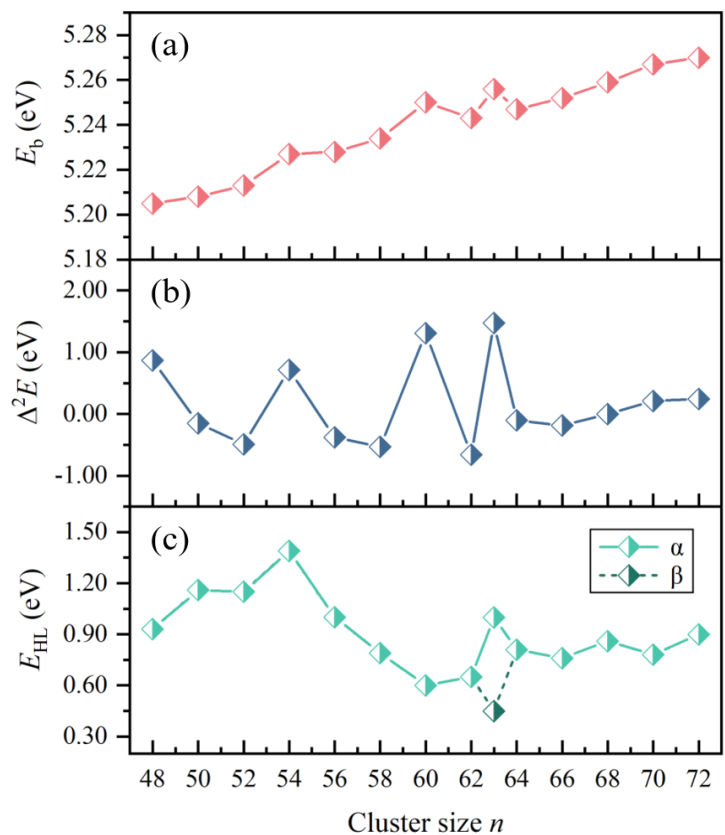

**Fig. 2** (a) Binding energy ($E_b$), (b) second-order energy differences ($\Delta^2 E$), and (c) HOMO-LUMO gap ($E_{HL}$) (split into α and β for $B_{63}$ cluster) of $B_n$ ($n$ = 48−72) clusters as a function of cluster size.

To gain a better understanding of the electronic structure of the bilayer pattern, ELF and AdNDP bonding analyses of the $B_{63}$ (I) cluster were conducted, as illustrated in Fig. 3. The $ELF_\pi$ results indicated the formation of delocalized π bonds along three $B_3$ chains at the waist, while the $ELF_\sigma$ pattern indicated the formation of localized B–B σ-bonds on the periphery and between the top and bottom layers, as well as delocalized in-plane σ bonds among the inner boron atoms. This bonding picture was consistent with the in-plane color-filled map illustrated in Fig. 4a. The high ELF distributions (close to 1) in interlayer bonds between bilayers demonstrated strong covalent bonding, and the delocalized distribution of the three $B_3$ chains on the waist was in agreement with the iso-surface characteristics of $ELF_\pi$.

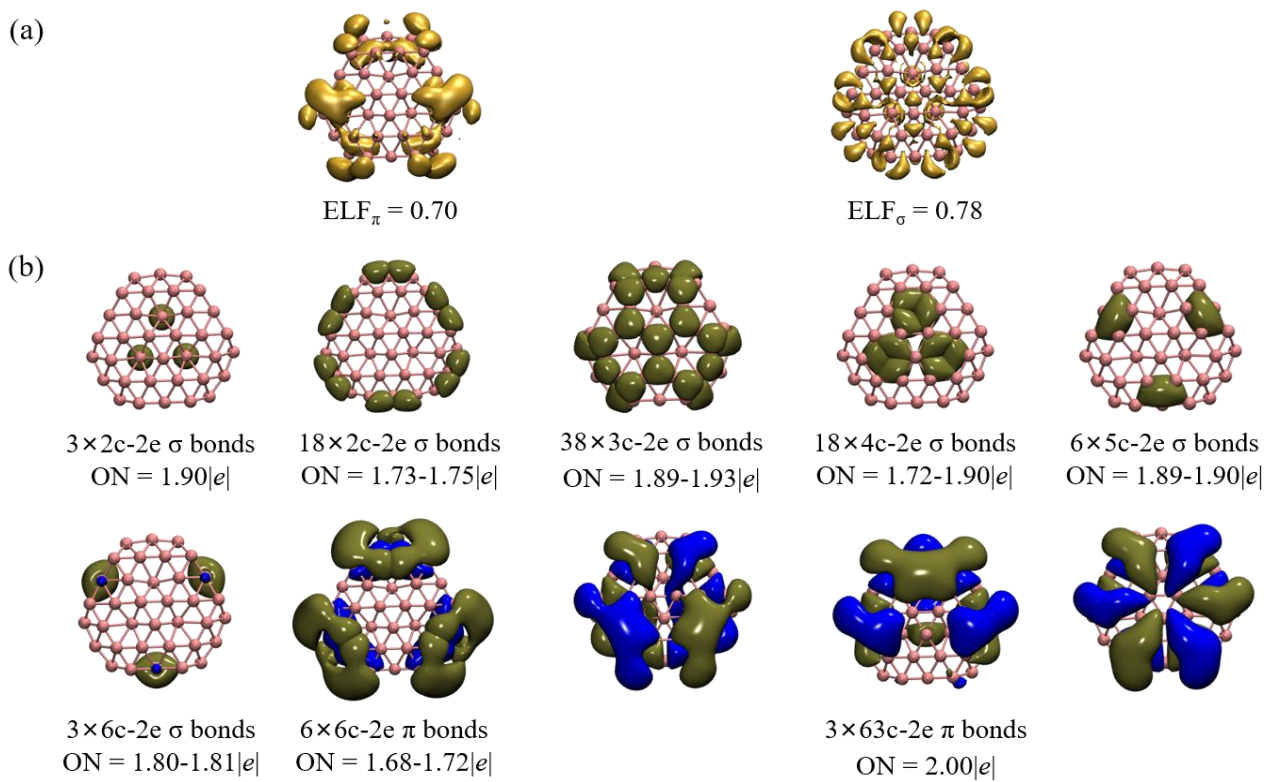

**Fig. 3** (a) The ELF analysis of $B_{63}$ (I) at the (U)TPSSh/6-311+G(d) level and (b) the AdNDP analysis of closed-shell anionic $B_{63}$ (I) at the TPSSh/6-311+G(d) level.

The AdNDP bonding analysis of the optimized closed-shell anionic $B_{63}$ (I) was conducted to assess its structural stability (Fig. 3b). This analysis revealed three 2c-2e σ bonds shared by six inward-buckled atoms bridging two layers, three 6c-2e σ bonds on lateral windows, eighteen 2c-2e σ bonds and six 3c-2e σ bonds on the periphery linking the top and bottom layers, as well as thirty-two 3c-2e σ bonds, eighteen 4c-2e σ bonds and six 5c-2e σ bonds covering both the top and bottom layers, which were consistent with the ELF results. Additionally, the remaining nine electron pairs formed a π-bonding system across the bilayer surface, including six delocalized 6c-2e π bonds along each $B_3$ chain in accordance with the electronic delocalization of the $ELF_\pi$ results and three completely delocalized 63c-2e π bonds across the entire bilayer. This π system adhered to the $4N + 2$ Hückel rule ($N = 4$) of aromaticity. The neutral $B_{63}$ (I) exhibits π aromaticity comparable to its anionic $D_{3h}$ counterparts, as evidenced by the similarity of their occupied molecular orbitals (Fig. S5). This π aromaticity plays a role in the remarkable stability of the neutral $B_{63}$ (I). Furthermore, the ADCH charge distribution of $B_{63}$ (I) (Fig. 4b) reveals that electrons accumulate on the inward-buckled atoms with negative charges of −0.27 *e*, forming localized σ bonds and strengthening interlayer bonding, thereby further stabilizing the bilayer structure.

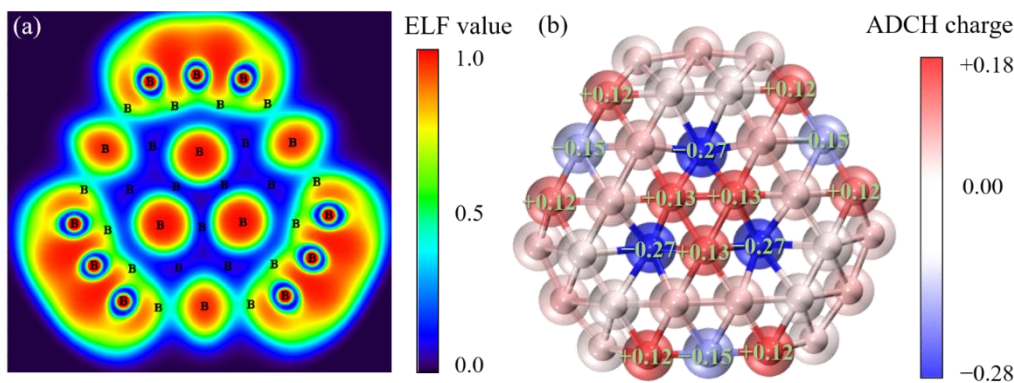

**Fig. 4** (a) the total ELF color-filled map in the central XY plane, and (b) ADCH charge distribution (in unit of *e*) of $B_{63}$ (I) with the charges varying from negative to positive, depicted by colors from blue to red.

To obtain further insight into aromatic properties of $B_{63}$ (I) cluster, we calculated the NICS(1) color-filled plane map at a distance of 1 Å from the center (Fig. 5a) and the scanned curves of NICS values in two different directions (Fig. 5b-5c). The results showed that cluster exhibited prototypical aromaticity, with the central six inward-buckled atoms displaying the most negative NICS(1) values (approximated −50 ppm). Furthermore, the ZZ component of the NICS(1) value was calculated to be −31.52 ppm, indicating a significant π aromaticity of $B_{63}$ (I). The NICS-scan curves of $B_{63}$ (I) along the y-axis and z-axis are indicative of a highly aromatic system, particularly due to the strong aromatic character of the central interlayer B−B bonds, which has a NICS value of approximately –60 ppm (Fig. 5b). Additionally, the upper and lower layers have NICS values of approximately –45 ppm (Fig. 5c). Furthermore, the NICS-scan curve in Fig. 5b displays two maxima along the y-axis at ± 5 Å with relatively low positive NICS values, indicating a weakly antiaromatic system, which is in line with the magnetic de-shielding effect distributed on the waist of the bilayer structure (Fig. S6). Meanwhile, the NICS-scan curve in Fig. 5c is symmetric with the negative NICS values decreasing rapidly with respect to the distance R from the bilayer center along the z-axis, which is in agreement with the magnetic shielding effect of the bilayer structure, thus indicating a three-dimensional aromatic structure.

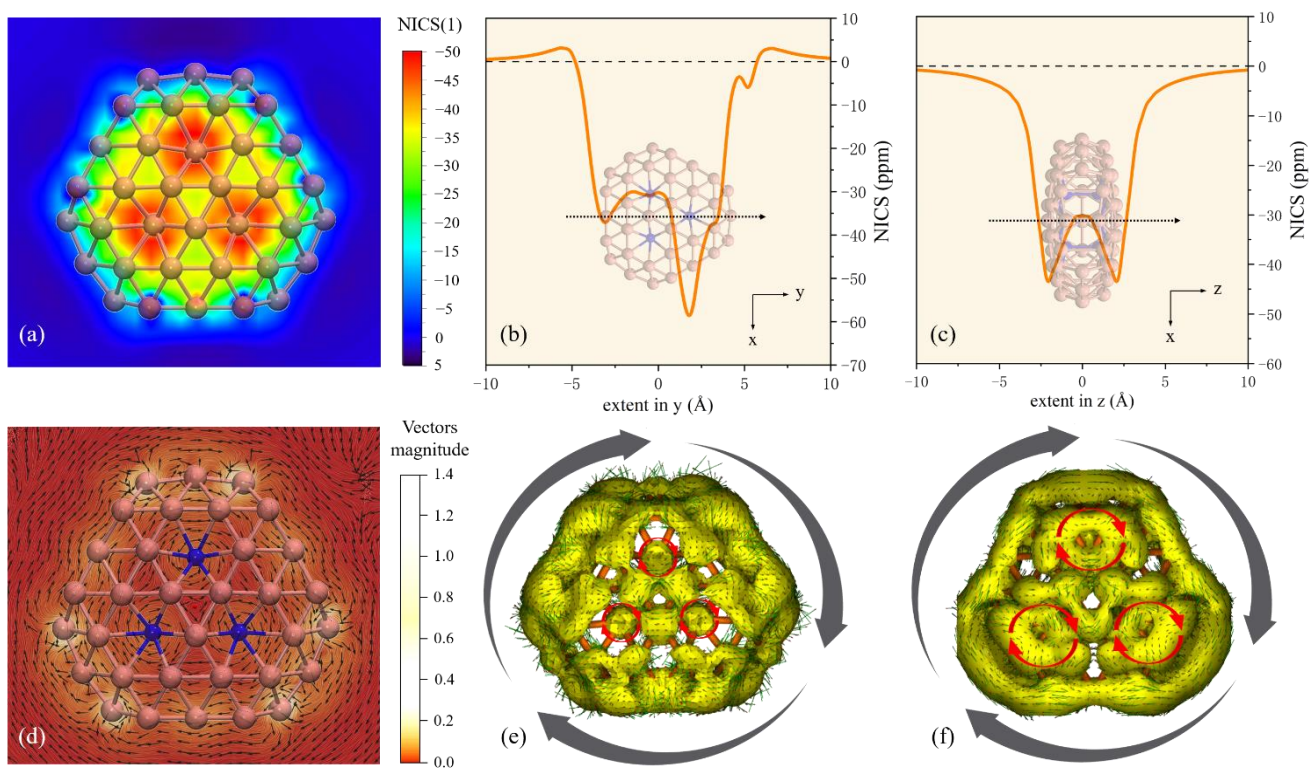

**Fig. 5** (a) the NICS(1) color-filled plane map above the central XY plane for 1 Å. The NICS-scanned curves on (b) the y-axis and (c) the z-axis of $B_{63}$ (I), respectively. (d) the GIMIC map of $B_{63}$ (I) on the central XY plane. (e-f) the ACID maps of $B_{63}$ (I) at different states (contributions from σ and π electrons). The external magnetic field is perpendicular to the ring plane and points outward with arrows indicating the direction of induced current, and diatropic currents corresponding to clockwise circulation of electrons, paratropic currents to anticlockwise circulation.

The magnetic responses of $B_{63}$ (I) compound to an external magnetic field were evaluated through the use of gauge-including magnetically induced currents (GIMIC)[50] map (Fig. 5d) and the ring current strength profiles along the plane, as well as the anisotropy of the current induced density (ACID) map's global diamagnetic ring current (Fig. 5e and 5f). The isosurfaces of the current density plot demonstrate two types of diatropic rings surrounding three local interlayer B−B bonds (8.0 nAT$^{-1}$) and encompassing the outer part of the cluster (35.0 nAT$^{-1}$). This magnitude is especially noteworthy when compared to the net ring current strength of 14.2 nAT$^{-1}$ for $C_6H_6$ at the same level, which implies an aromatic character attributed to the contributions of three local rings pivoting on interlayer B−B bonds and an external delocalized ring. Furthermore, the ACID isosurfaces of $B_{63}$ (I) and associated vectors of the induced current by applying an external magnetic field were divided into σ and π electron current density maps. It is evident from the ACID plots of $B_{63}$ (I) that both σ and π electrons were magnetically active, with significant current densities being detected. The outer loop ACID isosurface had a clockwise current circuit, indicating an aromatic character, while the inner loop also exhibited a clockwise current circuit, indicating electron delocalization and a certain double aromatic character. The analysis of GIMIC and ACID maps suggests that $B_{63}$ (I) compound has a double aromatic character, with both σ and π electrons exhibiting magnetic activity.

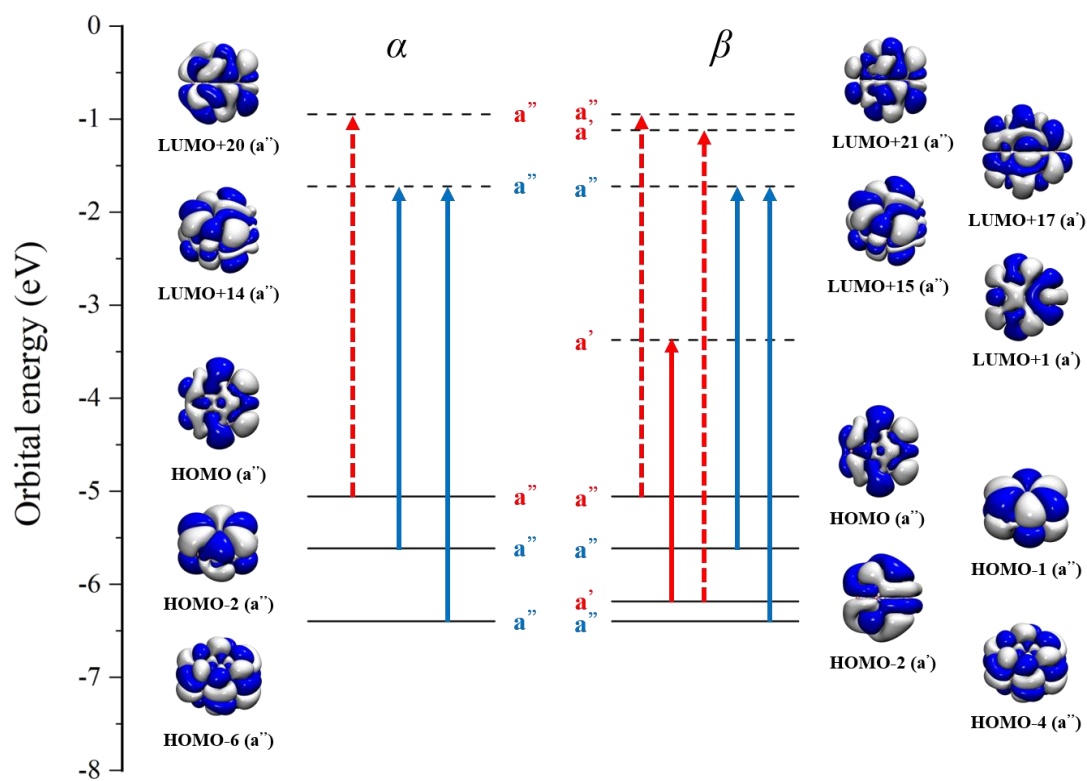

**Fig. 6** The frontier orbital energy levels (in eV) of α and β electrons of $B_{63}$ (I) calculated at the (U)TPSSh/6-311+G(d) level. The red (σ) and blue (π) arrows indicate transitions, with full arrows representing major translational transitions and dashed arrows representing rotational transitions.

The double aromatic character can be elucidated by analyzing the excitations responsible for the magnetic response, as illustrated in Fig. 6. According to the CTOCD-DZ method, the diamagnetic current density can be attributed to virtual translational transitions, while the paramagnetic current density is attributed to rotational transitions. The magnitude of a given occupied-unoccupied virtual transition is determined by the value of the respective linear and angular momentum matrix elements and the energy difference between the two orbitals.[52] As the energy gap between the occupied and unoccupied orbitals increases, the contribution of the transition diminishes. Therefore, the induced current density can be attributed to the virtual excitation of a few electrons in frontier orbitals.

Taking the Jahn-Teller distortion into account, we considered seven electrons (three σ and four π) in the doublet state of $B_{63}$ (I) with $C_s$ symmetry, which consists of a closed-shell π electronic configuration of (a'')$^2$ (a'')$^2$ and an opened-shell σ electronic configuration of (a'')$^2$ (a'')$^1$. The diatropic current density in the σ-electron subsystem was mainly derived from the translational transition between the HOMO-2 and LUMO+1 of $β$ electrons (a' × a' × a' = a'), while the paratropic currents generated from rotational transitions with larger orbital energy differences, slightly compensated the contribution of diamagnetic ring currents to aromaticity. In the π-electron subsystem, the translational operation was enabled by the $α$-electron transitions HOMO-2 and HOMO-6 to LUMO+14, as well as $β$-electron transitions HOMO-1 and HOMO-4 to LUMO+15 (a'' × a'' × a' = a'), which coincided with the global diamagnetic ring current in ACID maps. This illustration effectively demonstrates the diatropic currents observed at the σ and π electrons, thereby confirming the dual σ and π aromaticity of $B_{63}$ (I).

**Conclusions**

In summary, we present an extensive computational investigation of the structure, bonding, and stability of the $B_{63}$ cluster. Our findings indicate that this cluster is composed of a $C_s$ bilayer structure with three interlayer B−B bonds, the lengths of which are very close to those of bilayer borophene. We further observe that the $B_{63}$ (I) cluster is the most stable bilayer structure among the $B_{48}$−$B_{72}$ series. Through chemical bonding analyses, including ELF, AdNDP, and ADCH methods, we find that the bilayer structure features strong interlayer σ and π bonding interactions in the ground state with electrons accumulating on the inward-buckled atoms. More importantly, this bilayer complex also displays a strong aromatic character, primarily attributed to the local contributions of the three interlayer B−B bonds, and exhibits open-shell dual aromaticity. Our investigation verified the potential of an extending mode to effect the transformation of the most dependable $B_{63}$ bilayer into borophene nanomaterials.

**Acknowledgements**

This were supported by the National Natural Science Foundation of China (12104393 and 11804076), the Innovation Capability Improvement Project of Hebei province (NO. 22567605H).